\documentclass[twocolumn,showpacs,amsmath,amssymb,prl,floatfix]{revtex4}

\usepackage{graphicx}
\usepackage{dcolumn}
\usepackage{bm}

\begin{document}

\title{Two-band Effects in the Angular Dependence of $H_{c2}$ of
MgB$_{2}$ Single Crystals}

\author{A. Rydh}\email{rydh@anl.gov}
\author{U. Welp}
\author{A. E. Koshelev}
\author{W. K. Kwok}
\author{G. W. Crabtree}
\affiliation{%
Materials Science Division, Argonne National Laboratory, 9700 S. Cass 
Ave., Argonne, IL 60439, USA}
\author{R. Brusetti}
\author{L. Lyard}
\author{T. Klein}
\affiliation{%
Laboratoire d'Etudes des Propri{\'{e}}t{\'{e}}s Electroniques
des Solides, CNRS, BP 166, 38042 
Grenoble, France}
\author{C. Marcenat}
\affiliation{%
D{\'{e}}partement de Recherche Fondamentale sur la Mati{\`{e}}re
Condens{\'{e}}e, SPSMS, CEA-Grenoble, 38054 Grenoble, France}
\author{B. Kang}
\author{K. H. Kim}
\author{K. H. P. Kim}
\author{H.-S. Lee}
\author{S.-I. Lee}
\affiliation{%
NCRICS and Dept. of Physics, Pohang University of Science and 
Technology, Pohang 790-784, Republic of Korea}

\date{\today}

\begin{abstract}
The angular dependence of the upper critical field $H_{c2}$ of
MgB$_{2}$ single crystals is studied at various temperatures by means
of specific heat and transport measurements in magnetic fields up to
17~T. Clear deviations from Ginzburg--Landau behavior are observed at
all temperatures and are explained by two-band effects.  The angular-
and temperature dependence of the deviations are in qualitative
agreement with theoretical predictions based on band-structure
calculations.  Quantitative agreement is obtained with an interband
coupling slightly stronger than the calculated, enabling band-structure
anisotropies and interband coupling strength to be
experimentally estimated.  This provides a new pathway to the study of
disorder and doping effects in MgB$_{2}$.
\end{abstract}


\maketitle


The emergence of new theoretical works with close experimental
connections has significantly deepened the understanding of the
properties of magnesium diboride (MgB$_{2}$).  Despite the fact that
the superconducting properties of MgB$_{2}$ with its fairly simple
atomic structure were just recently discovered \cite{Nagamatsu01},
this phonon mediated s-wave superconductor has already been the
subject of intense and numerous studies \cite{ReviewPhC} due to its
exotic properties arising from a complex, disconnected, multi-band
Fermi surface.  Band-structure calculations have demonstrated that the
Fermi surface is composed of pairs of three-dimensional $\pi$-bands
and quasi-2D $\sigma$-bands \cite{Kortus01}.  This effective two-band
structure has been confirmed by de\,Haas\/--\/van\,Alphen
measure\-ments \cite{Yelland02} and angle-resolved photo\-emission
spectroscopy \cite{Uchiyama02}.

The superconducting properties of the two sets of bands are quite
different, due to the low overlap of the orthogonal $\sigma$- and
$\pi$-band wave functions.  The superconducting gap ranges from 1.5 to
3.5~meV on the $\pi$-bands and from 5.5 to 8~meV on the strongly
superconducting $\sigma$-bands \cite{Choi02}.  This double-gap nature
has been verified by tunneling experiments
\cite{Giubileo01,Iavarone02}, heat capacity measurements
\cite{Bouquet}, and spectroscopy \cite{Chen01,Szabo01,Schmidt02}.

Theoretically, two-band superconductivity has a history starting well
before MgB$_{2}$ \cite{Suhl59,Golubov97,Shulga98}.  Through
theoretical advances, a fairly unified picture has emerged with
predictions that can be experimentally substantiated
\cite{Choi02,Liu01,Brinkman02,Kogan02,Miranovic03,Gurevich03,Dahm03,
Golub-lam,Golub-band,Koshelev,Golubov0303237}.  One of the salient
predictions associated with a pronounced two-band effect is a
difference between the coherence length anisotropy
$\gamma_{\xi}=\xi_{ab}/\xi_{c}$
\cite{Miranovic03,Gurevich03,Dahm03,Golubov0303237} and the
penetration depth anisotropy
$\gamma_{\lambda}=\lambda_{c}/\lambda_{ab}$ \cite{Kogan02,Golub-lam},
both of which become temperature dependent with opposite tendencies. 
For MgB$_{2}$, a strong decrease of
$\gamma_{\xi}=H_{c2}^{ab}/H_{c2}^{c}$ from $\gamma_{\xi}(0) \sim 5$ to
$\gamma_{\xi}(T_{c})\approx 2$ is found experimentally
\cite{Angst02,Welp03,Lyard02,Machida03,Zehetmayer02,Cubitt03}, while
controversy remains about the experimental temperature dependence of
$\gamma_{\lambda}$ \cite{Zehetmayer02,Cubitt03, LyardHc1}.

In this Letter we present evidence of clear deviations of the angular
dependence of $H_{c2}$ from the anisotropic Ginzburg--Landau (GL)
description.  The $H_{c2}(T,\theta)$ transition of MgB$_{2}$ single
crystals was determined from resistivity measurements and specific
heat with excellent agreement between the two.  With a slight
adjustment of some of the parameters supplied by band-structure
calculations, good quantitative agreement is found between theory
\cite{Golubov0303237} and experiment, yielding fundamental estimates
of band-structure anisotropies and the interband coupling strength.

Several MgB$_{2}$ crystals with typical dimensions 50 -- 250~$\mu$m
were obtained through a high pressure heat treatment of a mixture of
Mg and B in excess Mg as described elsewhere \cite{Jung02}.  The
crystals had $T_{c}$ values of 34 -- 36~K and a $H_{c2}^{c}(0) \approx
3.5$~T. Transport measurements were performed using standard AC
techniques at 23~Hz with a resolution better than 0.5~nV. For specific
heat measurements, the crystals were mounted on top of flattened
12.7~$\mu$m chromel/constantan thermocouple junctions.  Small
temperature oscillations of the sample were induced by either a
resistive heater wire (Sample~1, Argonne) or by modulating the
temperature of the copper base \cite{Graebner89} (Sample~2, Grenoble). 
The temperature oscillation was measured through the AC voltage across
the junction, the absolute base temperature monitored by a Cernox
thermometer, and the sample temperature offset obtained from the
DC voltage of the thermocouple.

\begin{figure}
\includegraphics[width=0.92\linewidth]{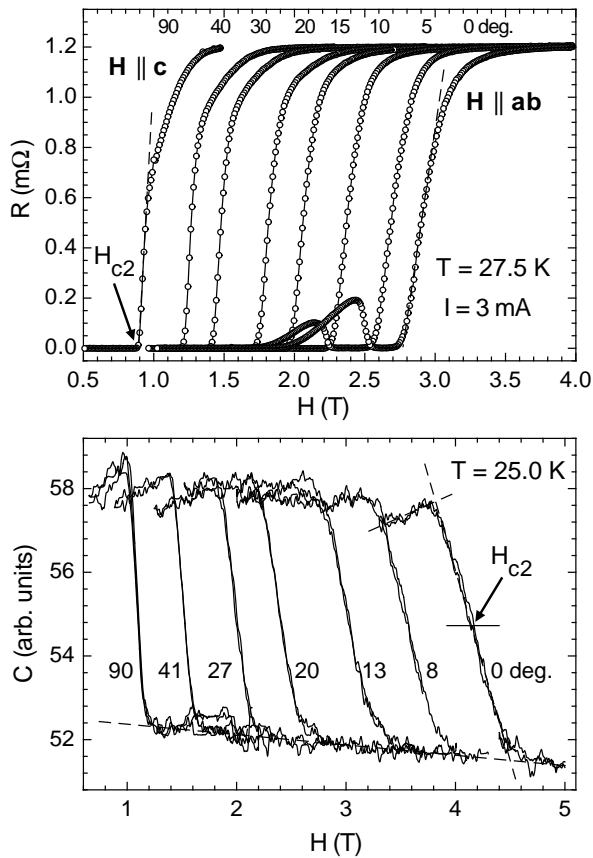}
\caption{Top: Resistive $H_{c2}$ transition at $T=27.5$~K as a
function of magnetic field $H$ and angle $\theta$ from the basal
plane.  Bottom: Corresponding specific heat signature at $T =
25.0$~K.}
\label{Fig1}
\end{figure}

Figure~\ref{Fig1} shows the transitions from resistivity (top) and
specific heat (bottom) as a function of angle at $T=27.5$~K and $T =
25.0$~K, respectively.  The resistive transitions were measured at a
relatively high current density to suppress the effects of surface
superconductivity at the well-shaped crystal surfaces, as discussed in
\cite{Rydh03}.
The value of $H_{c2}(\theta)$ was determined through a linear
extrapolation of the steep drop to zero resistivity as shown by the
nearly vertical dashed lines in the top panel.  The appearance of the
peak effect just below $H_{c2}$ for some angles is evident in the
figure.  The thermodynamic signature of $H_{c2}$ was defined from the
midpoint of the specific heat transitions, as illustrated in the
bottom panel.  The choice of definition was checked not to be
significant.  It is interesting to note that the specific heat step
height is fairly independent of the field direction.  This is in
agreement with GL theory, where the step height should scale with
$T(\mathrm{d}H_{c}/\mathrm{d}T)^{2}$, where $H_{c}$ is the (isotropic)
thermodynamic critical field.  Possible deviations from a constant
step height arising from two-band effects are too small to be resolved
in the current data due to uncertainties in the experimental method.

\begin{figure}
\includegraphics[width=0.91\linewidth]{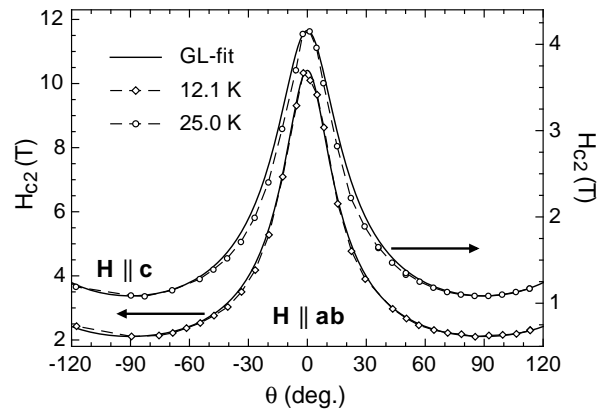}
\caption{Angular dependence of the upper critical field at 12.1~K and
25.0~K. The solid lines correspond to the GL theory.  Small, but clear
and consistent deviations from the anisotropic, effective-mass
description are seen.}
\label{Fig2}
\end{figure}

Clear deviations from an anisotropic GL description are, however, seen
in the angular dependence of $H_{c2}$.  In Fig.~\ref{Fig2}, the
$H_{c2}(\theta)$ curves are shown for two selected temperatures
together with corresponding fits to the effective mass
description $H_{c2}^{\mathrm{GL}}(\theta) =
H_{c2}^{ab}/(\cos^{2}\theta +\gamma_{\xi}^{2}\sin^{2}\theta)^{1/2}$.
The relative deviations are fairly small at 12.1~K as compared to
25.0~K. They are nevertheless clearly discernible at all temperatures
and are reproducible between different measuring setups, samples, and
methods.  Resistive measurements by Eltsev~{\it et\,al.} displayed
similar deviations but were not analyzed in detail \cite{Eltsev02}. 
Deviations were also reported at 33~K using torque measurements
\cite{Angst}.  The latter, however, suffer from the inability to
measure $H_{c2}$ along the symmetry axes.

\begin{figure}
\includegraphics[width=0.91\linewidth]{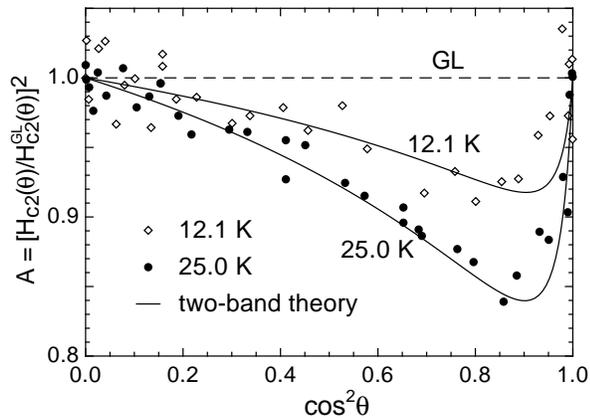}
\caption{Ratio $\mathcal{A}=[H_{c2}(\theta)/H_{c2}^{\mathrm{GL}}
(\theta)]^{2}$ as a function of $\cos^{2}\theta$ for the two
temperatures of Fig.~\ref{Fig2}.  The dashed line $\mathcal{A}=1$
corresponds to the GL description, assuming an anisotropy parameter
$\gamma_{\xi}=H_{c2}^{ab}/H_{c2}^{c}$ that is allowed to vary as a
function of temperature.  Solid curves are given by two-band theory
with best-fit parameters as discussed in the text.}
\label{Fig3}
\end{figure}

The deviations from GL behavior are illuminated by plotting the ratio
$\mathcal{A}=[H_{c2}(\theta)/H_{c2}^{\mathrm{GL}} (\theta)]^{2}$ as a
function of $\cos^{2}\theta$ as shown in Fig~\ref{Fig3}.
When the field is directed along the $c$ axis or within the basal
plane there are no deviations, since the experimental $H_{c2}^{ab}$
and $H_{c2}^{c}$ were used as parameters for the GL-fit
($\mathcal{A}=1$) at each temperature.  The shape of the deviations as
a function of angle is similar for all temperatures, with a maximum
amplitude at around $\theta = 20\/^{\circ}$--$30\/^{\circ}$, i.e., for
$\cos^{2}\theta \sim 0.9$.  The theoretical curves are discussed
below.

\begin{figure}
\includegraphics[width=0.90\linewidth]{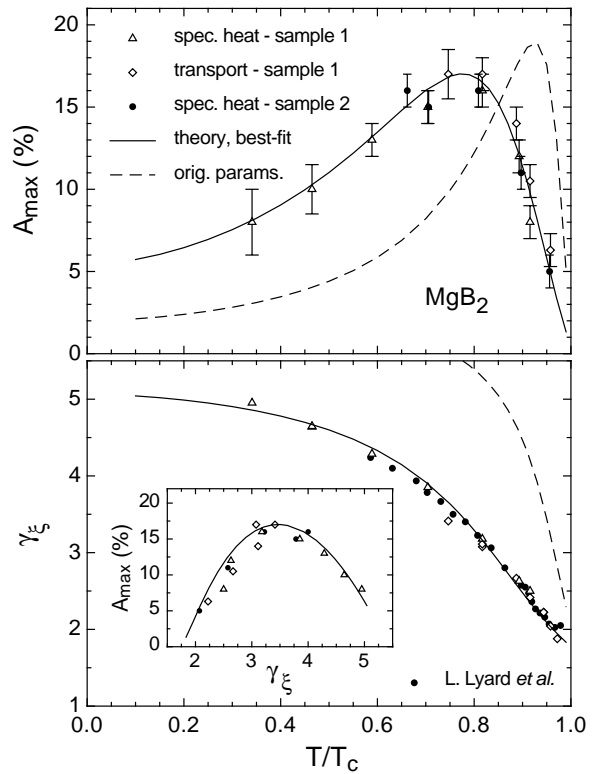}
\caption{Top: Maximum deviations of $\mathcal{A}(\theta)$ from the GL
theory as a function of reduced temperature.  The dashed curve is
taken from \cite{Golubov0303237}.  Bottom: Temperature dependence of
the experimental anisotropy $\gamma_{\xi}$.  Solid circles are taken
from from \cite{Lyard02}.  Other symbols are as above.  The inset
shows $\mathcal{A}_{\mathrm{max}}$ as a function of anisotropy,
illustrating maximum deviations at intermediate values of
$\gamma_{\xi}$ where both bands contribute equally.  Parameters for
the theoretical curves are given in Table~\ref{Table1}.}
\label{Fig4}
\end{figure}

To investigate the temperature dependence of the deviations in more
detail, the maximum amplitude of $\mathcal{A}(\theta)$ is plotted as a
function of temperature in Fig.~\ref{Fig4} (top).  Good agreement is
found between the transport and specific heat data from Argonne
(Sample~1, measured on the same crystal, $T_{c}= 35.5$~K) and specific
heat measurements taken at Grenoble (Sample~2, $T_{c}\approx 34$~K),
illustrating the fundamental and consistent nature of the deviations. 
The amplitude is relatively small at low temperatures and reaches a
maximum slightly below $T_{c}$.  By comparing the temperature
dependence of $\mathcal{A}_{\mathrm{max}}$ with that of $\gamma_{\xi}$
(bottom panel) one can see that the maximum of
$\mathcal{A}_{\mathrm{max}}(T)$ occurs at intermediate values of
$\gamma_{\xi}$.

The general experimental features of $H_{c2}(T,\theta)$ can be
excellently described by the recent theory of the angular dependence
of dirty two-band superconductors \cite{Golubov0303237}.  The two-band
theory requires as input (i) the matrix of effective coupling
constants
$\Lambda_{\alpha\beta}=\lambda_{\alpha\beta}-\mu_{\alpha\beta}^{\ast}$,
where $\lambda_{\alpha\beta}$ are the electron-phonon coupling
constants and $\mu_{\alpha\beta}^{\ast}$ are the Coulomb
pseudopotentials ($\alpha$ and $\beta$ are indices for the $\sigma$-
and $\pi$-bands), (ii) band anisotropies $\gamma_{\sigma}$ and
$\gamma_{\pi}$, and (iii) a ratio of the diffusion constants in the
two bands, e.g., $r_{z}=\mathcal{D}_{\pi,z}/\mathcal{D}_{\sigma,z}$. 
The theoretical dependencies of $\mathcal{A}(T)$ and $\gamma_{\xi}(T)$
obtained by using coupling constants and anisotropies supplied by
band-structure calculations \cite{Golub-band,Brinkman02} are
illustrated by the dashed curves in Fig.~\ref{Fig4}.  It is clear
that, while qualitatively similar, the theoretical curves are
displaced closer to $T_{c}$ and the predicted anisotropy is higher
than the experimental one.

The shapes of the theoretical curves are sensitive to mainly two
parameters, the interband coupling strength, expressed through the
reduced parameter $S_{12}\approx\Lambda_{\sigma\pi}\Lambda_{\pi\sigma}
/{\left({\Lambda_{\sigma\sigma}-\Lambda_{\pi\pi}}\right)}^{2}$
\cite{Golubov0303237}, and the ratio $r_{z}$.  In particular, the
overall change of anisotropy can be estimated as
$\gamma_{\xi}(T_{c})/\gamma_{\xi}(0)\approx 1/\sqrt{1+S_{12}r_{z}}$. 
We found that a quantitative description of the experimental data
requires (i) a significantly increased relative interband coupling
$S_{12}$, (ii) a somewhat decreased anisotropy $\gamma_{\sigma}$ of
the $\sigma$-band, and (iii) an almost isotropic $\pi$-band.  $S_{12}$
was increased by augmenting the two off-diagonal coupling constants by
a factor $1.9$ from the values provided in
Ref.~[\onlinecite{Golub-band}].  The set of parameters that gives the
best description of the experimental data is listed in
Table~\ref{Table1} together with the original parameters.  The
resulting fits are shown as solid curves in Figs.~\ref{Fig3} and
\ref{Fig4}.  We note that the fits in Fig.~\ref{Fig4} allow for an 
independent determination of
$\mathcal{D}_{\sigma,x}/\mathcal{D}_{\pi,x}
\equiv {({\gamma_{\sigma}/\gamma_{\pi}})}^{2}/r_{z} \approx 0.23$, in
good agreement with the observation of an enlarged vortex core
\cite{Koshelev,Eskildsen02}.

A possible source of the discrepancy between the theoretical
calculations and the experiments is a theoretical overestimation of
the off-diagonal Coulomb pseudopotentials (see discussion in
Ref.~[\onlinecite{MazinPhysC03}]) resulting in too low values of
$\Lambda_{\sigma \pi}$ and $\Lambda_{\pi \sigma}$.  Unfortunately, no
direct experimental probe of the off-diagonal coupling constants is
available at present.  On the other hand, the $T_{c}$ values of our
single crystals are somewhat lower than for polycrystalline samples. 
Thus, it is also possible that the discrepancy is due to a slightly
modified band-structure arising from a non-stoichiometric composition
of the crystals.  Another origin of $T_{c}$ suppression in single
crystals could be interband impurity scattering.  A natural question
is how interband scattering affects the anisotropic properties.  By
analyzing the theoretical corrections to the components of the upper
critical fields from the inclusion of weak interband scattering, we
conclude that it is unlikely that this scattering is responsible for
discrepancies between the calculated band-structure and experiment
(e.g., for the lower value of $\gamma_{\sigma}$)
\footnote{In the main order with respect to the interband scattering
rate $\Gamma_{\sigma\pi}$, interband impurity scattering acts as a
pair breaker, homogeneously suppressing $T_{c}$ and both components of
$H_{c2}$ without modifying the anisotropy.  The leading corrections to
the anisotropy have the order
$\Gamma_{\sigma\pi}\Lambda_{\pi\sigma}/\Lambda_{\sigma\sigma}$ meaning
that the relative change of anisotropy due to the interband scattering
at any temperature has to be significantly smaller than the relative
change of $T_{c}$.  In addition, the correction to the basal-plane
upper critical field has an extra small factor $1/\sqrt{r_{z}}$.}.

\begin{table}
\caption{Parameters used in the theoretical computations.}
\label{Table1}
\begin{ruledtabular}
\begin{tabular}{ccc}
Parameter&\mbox{Predicted Params.\footnote{As predicted by
band-structure calculations.  $\Lambda$-values are taken from
Ref.~[\onlinecite{Golub-band}] and $\gamma$-values are obtained from
Ref.~[\onlinecite{Brinkman02}].  The parameter $r_{z}=300$ was not 
calculated but estimated, and corresponds to
$\mathcal{D}_{\sigma,x}/\mathcal{D}_{\pi,x} \equiv
{({\gamma_{\sigma}/\gamma_{\pi}})}^{2}/r_{z} = 0.2$
\cite{Koshelev}.}}&\mbox{Best Experimental Fit}\\
\hline
$\gamma_{\pi}$ & 0.82 & {\bf 1.02}\\
$\gamma_{\sigma}$ & 6.3 & {\bf 5.4}\\
$S_{12}$ & 0.034 & 0.105\\[0.2mm]
$\left({\begin{array}{cc}
    \Lambda_{\sigma \sigma} & \Lambda_{\sigma \pi}  \\
    \Lambda_{\pi \sigma} & \Lambda_{\pi \pi}
\end{array}}\right)$
&
$\left({\begin{array}{cc}
    0.81 & 0.115  \\
    0.091 & 0.278
\end{array}}\right)$
&
$\left({\begin{array}{cc}
    0.81 & {\bf 0.216} \\
    {\bf 0.171} & 0.278
\end{array}}\right)$
\\[1.7mm]
$r_{z}=\mathcal{D}_{\pi,z}/\mathcal{D}_{\sigma,z}$ & 300 & 120\\
\end{tabular}
\end{ruledtabular}
\end{table}

MgB$_{2}$ single crystals are usually described as fairly clean, with
the $\sigma$-band probably in the clean limit
\cite{Yelland02,Eskildsen02,Quilty03}.  This is supported by the low,
reproducible value of $H_{c2}^{c}(0) \approx 3.5$~T. The temperature
dependence of $\gamma_{\xi}$ has also been described successfully
within the clean-limit formalism \cite{Dahm03}.  However, this still
requires similar modifications of coupling and band anisotropies from
the predicted values.  To our knowledge, a clean-limit calculation of
the angular dependence of $H_{c2}$ has not yet been presented.  One
could expect deviations from the anisotropic GL dependence of $H_{c2}$
at low temperatures even for a clean $\sigma$-band due to Fermi
surface effects.  On the other hand, the theory \cite{Golubov0303237}
for $H_{c2}(\theta)$ should remain valid even for a clean
$\sigma$-band in the vicinity of $T_{c}$, where this band is described
by GL theory and only the contribution from the $\pi$-band requires a
microscopic approach.

In summary, we have studied the angular and temperature dependence of
the upper critical field of MgB$_{2}$ single crystals by means of heat
capacity (specific heat) and transport measurements.  Clear two-band
effects are found in both $H_{c2}(\theta)$ and the temperature
dependence of the upper critical field anisotropy $\gamma_{\xi}(T)$. 
The experiments are well explained by the theory
\cite{Golubov0303237}, providing a deep understanding of the
microscopic parameters describing the system.  This work thus points
out a new pathway to the study of disorder and doping effects in
MgB$_{2}$, with great implications for future applications.

Support was provided through the Fulbright program and the
Sweden-America Foundation (A.R.), by the Ministry of Science and
Technology of Korea, and by the U.S. Department of Energy, Basic
Energy Sciences, under Contract No.~W-31-109-ENG-38.  We thank A. A.
Golubov for discussions and S. Hannahs, T. Murphy, and E. Palm for
assistance with measurements at NHMFL.

\end{document}